# COVID-19 MODELING TOWARDS SOCIOECONOMIC AND HEALTH DATA FROM NEW SOUTH WALES (NSW) - AUSTRALIA: AN APPROACH VIA GEOSPATIAL ANALYSIS AND GEOGRAPHICALLY WEIGHTED POISSON REGRESSION (GWPR)

Francelino A. XAVIER-CONCEIÇÃO

(francelino.conceicao@gmail.com or fcon069@aucklanduni.ac.nz)

## Abstract

An integrated approach of spatial data analysis and Geographically Weighted Poisson Regression (GWPR) along with global regression techniques are used in this study. This approach aims to model relationships between dependent variable Covid-19 and independent variables from socioeconomic and pre-existing health conditions within the local government area (LGA) in New South Wales (NSW)-Australia. Based on geospatial data analysis and a step-by-step procedure in building both global and GWPR models, four (4) independent variables are finally selected to investigate relationships between dependent and independent variables at the local scale. The GWPR model's results with the Goodness-of-Fit ($R^2$) range between 45-73% exhibit positive relationships between Covid-19 and the total population, the cancers, and the people with ages between 60 and 85 in most of the NSW state. Meanwhile, a negative relationship is observed between Covid-19 and the ischaemic heart disease; however, the estimated coefficients for this relationship are very low and close to zero; hence further investigation, including assessment from a different perspective, is necessary for validation. In conclusion, the model suggests that the relationships between the dependent variable and independent variables are nonstationary. Therefore, GWPR model calibration plays a vital role in geographic modelling at the local scale.

## 1. Introduction

Covid-19 is "an infectious disease caused by a newly discovered coronavirus" (WHO, 2020), which was first reported from China on December 31st, 2019 (WHO, 2020). Since then, the pandemic has spread rapidly throughout the world and resulted in a total of 8,525,042 active cases and 456,973 deaths by June 20th, 2020. The total infected cases include 3,144 positive cases and 48 deaths reported from New South Wales, Australia (Department of Health, Australian Government, 2020). Furthermore, WHO (2020) declared that older people, including those with pre-medical health problems, such as cardiovascular disease, diabetes, chronic respiratory disease, and cancer, are more likely to be infected with severe illness.

In this study, the author statistically analyses the spatial data, namely independent variables, and geographically models the Covid-19, namely dependent variable, to investigate the relationship between the dependent and independent variables by the local government area (LGA) in New South Wales (NSW), Australia. The global regression and the Geographically Weighted Poisson Regression (GWPR) approaches are used to model the relationship globally and geographically between Covid-19 and socioeconomic variables as well as the pre-existing medical health problems, particularly those susceptible to coronavirus as declared WHO (2020).

## 2. Spatial data and Methodology

### 2.1 Study Area and Spatial Data

The study area, New South Wales (NSW) is geographically located on the east coast of the Australian continent with a total area of 809,444 km². NSW state shares onshore boundaries with nearby states, Queensland in the north, South Australia in the West, and Victoria in the south. Meanwhile, its eastern edge directly facing the Tasman Sea of the Pacific Ocean. The NSW state's location is illustrated in figure 1, along with the total Covid-19 cases in each LGA between January 25th, 2020 and April 24th, 2020.

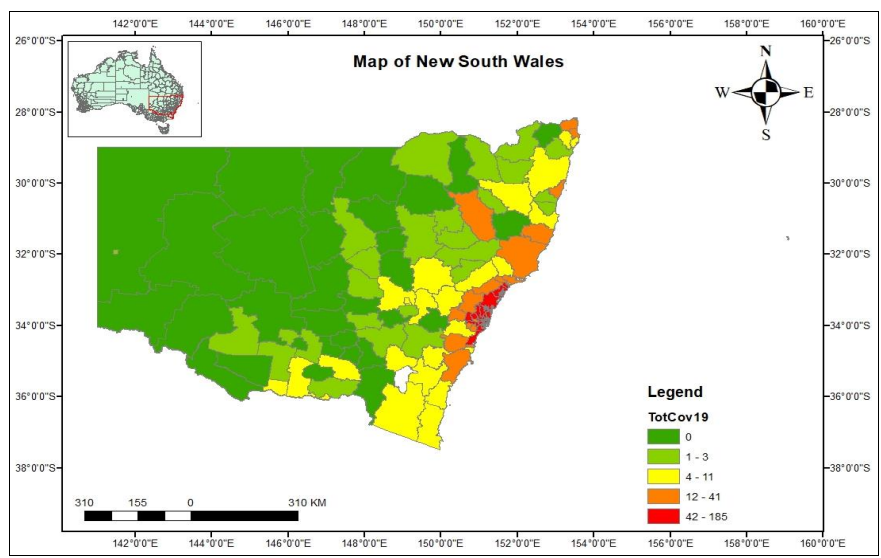

*Figure 1. Map of New South Wales and total Covid-19. The different colours indicate total infected cases per LGA by April 24th, 2020.*

Dataset used in this study primarily consists of two main groups, the socioeconomic data and the health records in New South Wales between 2016 and 2017, which were obtained from *Data.NSW*, *Public Health Information Development Unit (PHIDU)*, and *Australian Bureau of Statistics*. The dependent variable, the total cases of Covid-19 in New South Wales, were obtained from *Data.NSW* (https://data.nsw.gov.au/data/dataset/covid-19-cases-by-location/resource/21304414-1ff1-4243-a5d2-f52778048b29). The Covid-19 data is available on a day-to-day basis for each LGA between January 25$^{th}$ and April 24$^{th}$, 2020. This data is then re-arranged into total Covid-19 cases for each 130 LGA in the New South Wales state. Meanwhile, the socioeconomic data was primarily obtained from the Australian Bureau of Statistics (https://www.abs.gov.au/AUSSTATS/abs@.nsf/DetailsPage/3105.0.65.0012016?OpenDocument), and partially obtained from PHIDU (http://phidu.torrens.edu.au/social-health-atlases/data) along with the data for pre-existing medical health problems. A list of the dataset used in this study is provided in Annex 1.

## 2.2 Global Regression Linear for global/aspatial model

Global regression linear is a standard stationary regression linear for an aspatial model (Fotheringham et al., 2013). The relationship between the dependent and independent variables is expressed by a constant mean value across the study area regardless of variability over space. A standard formula for the global regression calibration is expressed as follows:

$$Y_i = \beta_0 + \beta_1 X_{i1} + \beta_2 X_{i2} + \ldots \beta_n X_{in} \qquad \text{(Equation 1)}$$

Where $Y_i$ is the dependent variable (in this case total Covid-19 cases) observed at location i, $X_i$ represents independent variables measured at location i, and β represents coefficient parameters that describe how a change in X (independent variable) affects Y (dependent variable).

In relation to geographically weighted regression linear discussed in the later stage, the global model and its resulting parameters are important as a benchmark model to be compared to its counterpart GWPR local model (Fotheringham et al., 2003). In terms of operation of the GWR4 software, the global model is generated simultaneously with the GWPR model.

## 2.3 Geographically Weighted Regression Linear for GWPR model

Unlike the global model, the geographically weighted regression linear is a nonstationary regression linear with an additional function of spatial location (Fotheringham et al., 2003), where the estimated coefficient parameters are varied over space. A formula for the GWPR regression is expressed with variability over space as follows:

$$Y_i = \beta_0(i) + \beta_1(i) X_{i1} + \beta_2(i) X_{i2} + \ldots \beta_n(i) X_{in} \qquad \text{(Equation 2)}$$

where index i is added to the standard global regression calibration formula in equation 1 to express the variability over space locally.

In this study, a step-by-step procedure in building the GWPR model is adopted from Fotheringham et al. (2003), Vanessa da Silva (2015), Tenerelli et al. (2016), and Mansley et al. (2015); and provided in the following 6 (six) stages in order.

2.3.1 Stage 1: Variable selection

The GWPR modelling is initiated by variable selection in order to appropriately select the independent variables based on the literature review, the recommendation from experts, and the common knowledge with consideration to the dependent variable. In this section, a total of 30 independent variables are selected from 50 variables from the dataset provided in annex 1. The independent variables selected consist of 5 variables from the socioeconomic group and 25 variables from the health group.

2.3.2 Stage 2: Multicollinearity check

Multicollinearity between variables is assessed using "Spearman's correlation rank" with the threshold correlation of 0.7 (Tenerelli et al., 2016). This process is aimed to filter out variables with the multicollinearity effect due to the high degree of correlation between variables. At this stage, variables with a correlation beyond 0.7 are excluded except those considered susceptive to coronavirus infection, such as older people, cardiovascular disease, diabetes, chronic respiratory disease, and cancers, as indicated by WHO (2020). Therefore, a total of 18 variables were further rejected and living 12 variables to proceed to the next stage, which consists of 3 socioeconomic variables and 9 health variables. The multicollinearity check's result is presented as a correlation matrix between independent variables as provided in annex 2.

2.3.3 Stage 3: Model optimization with a stepwise-AICc procedure

Akaike Information Criterion (AICc) is a measurement of statistical model's quality from a dataset (Konishi et al., 2008). The quality optimization of the final model can be achieved by comparing AICc values of multiple models calibrated from the dataset; the lower the AICc, the better the model quality (Fotheringham et al., 2003). The stepwise-AICc procedure is initiated by sequentially run ordinary least square (OLS) for each variable from the dataset, the OLS round one. The variable with the best fitting model, which is the smallest AICc value, is selected and added permanently to the model before proceeding to the next rounds. This procedure is repeated until the smallest AICc value is obtained for all the independent variables from each round. The next step is to plot the smallest AICc values of each independent variable in the scatter plot and assess the quality optimization of the model from the curve. The variable with the highest AICc value from OLS round one declines towards the curve's lowest point before reaching its constant point or inclining point. The variables with AICc constant or inclining after the lowest point considered no improvement to the final model. This procedure can be simply run in RStudio programming software with appropriate programming code for statistical computation following the aforementioned procedure.

At this stage, a total of 12 variables from stage 2 are further assessed with the stepwise AICc procedure. It reveals that the last 5 variables no longer improve the model, and therefore they are excluded from the final model. Variables excluded are Mood-Disord, Obst_Pulmonary, SQKM, Diabetes and Stroke; hence leaving the 7 variables: TotPop, Rest_syst, Cancers, 60-85yr, Kidney, Ischaemic_heart and Asthma. The result of the stepwise AICc procedure can be seen in figure 2.

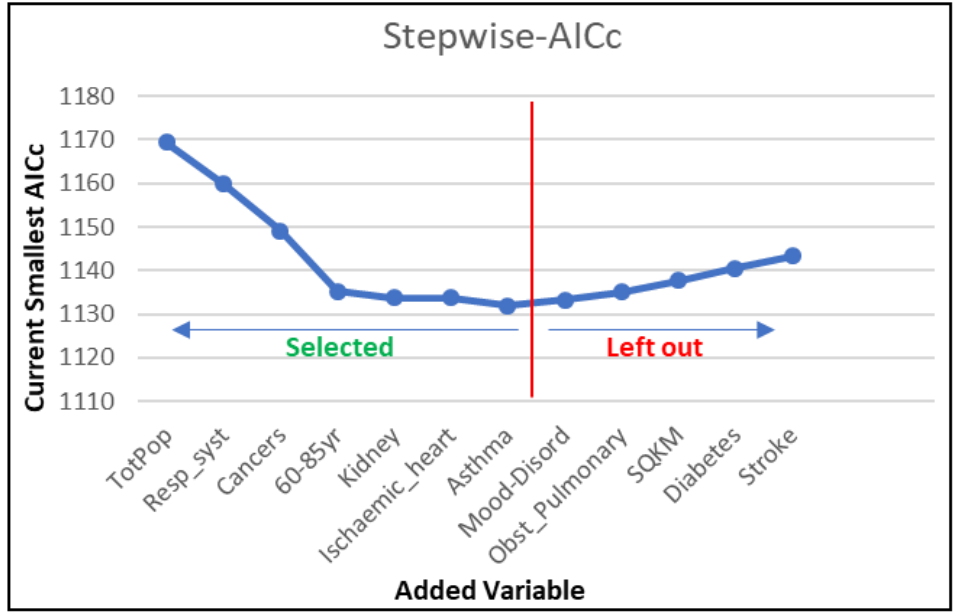

*Figure 2. Stepwise-AICc procedure for model quality optimization (variables Mood-Disord, Obst_Pulmonary, SQKM, Diabetes, and Stroke are excluded)*

2.3.4 Stage 4: Spatial variability test

The spatial variability of an individual variable can be tested by comparing the original GWR model and the switched model, the comparison indicators (Nakaya, 2016). In GWR4 software, the spatial variability is given by "Diff of Criterion," which indicates the comparison coefficient between the original model and the switched model. The variables with a positive "Diff of Criterion" value indicate no local spatial variability, while negative values indicate local spatial variability (Nakaya, 2016); hence, decisions can be made in terms of local and global variables. At this stage, the spatial variability test's results indicate that variables Resp_syst, Kidney, and Asthma are observed to have very low/no local spatial variability as given by the positive "Diff of Criterion" coefficient. Therefore, these variables are only assigned to the global model and excluded from the local model calibration. The result of the spatial variability test is provided in table 1.

*Table 1. Result of spatial variability test (negative "Diff of Criterion" indicates no spatial variability)*

| Variable | F | DOF for F test | DIFF of Criterion |
|---|---|---|---|
| Intercept | 0.750099 | 0.554 120.811 | 0.863945 |
| TotPop | 5.127172 | 0.209 120.811 | -0.652167 |
| Resp_syst | 1.538115 | 0.313 120.811 | 0.224372 |
| Cancers | 3.160455 | 0.132 120.811 | -0.134786 |
| 60-85yr | 8.089597 | 0.139 120.811 | -0.875815 |
| Kidney | 0.618454 | 0.012 120.811 | 0.020069 |
| Ischaemic_heart | 5.430100 | 0.074 120.811 | -0.254740 |
| Asthma | 0.355863 | 0.085 120.811 | 0.169835 |

2.3.5 Stage 5: Model Calibration

The model calibration for both global and GWPR are run simultaneously in GWR4 software with a 5 (five) steps procedure (Nakaya, 2016). Step 1 is project settings and data input, including the final review of the dataset before proceeding to model calibration. Step 2 is the most critical part of the model calibration, where variables are assigned as dependent or independent variables as well as local or global, settings for model type, and define the geographical location of the variables. In this study, variables TotPop, Cancers, 60-85yr, and Ischaemic_Health are selected for both global and local regression, while the other three variables are limited to global regression with consideration to procedure discussed in stages 1 - 4; and the Poisson model type is selected for the count data. Step 3 defines the Kernel function for geographical weighting to estimate local coefficients, bandwidth size, and model selection criteria necessary for finding the best bandwidth. In this study, the adaptive bi-square Kernel type is selected with the bandwidth selection method of golden section search and AICc selection criteria. Steps 4 and 5 are settings for output files and further executing and run calibration for the final model.

2.3.6 Stage 6: Mapping, analysis, and interpretation

Global and local models obtained from GWR4 are further mapped and analysed in the ArcGIS software, including minor calculations in Microsoft excel. This process includes analysis and interpretation of the individual model as well as model comparison. Details of mapping, analysis, and interpretation are described in sections 3 and 4.

In order to analyse the spatial autocorrelation, the standard residuals for global and local models are calculated and mapped. The standard residuals for both global and local models are calculated with reference to estimated coefficients and statistical parameters generated from the final model. The global regression coefficients are provided in table 2, while local regression coefficients are indicated by 'yhat'. Calculations of standard residuals for each model are provided in the following equations 3 to 6, respectively.

Global Residual = [TotCov19] - (2.067951 + 0.000016 * [TotPop] + 0.000369 * [Cancers] - 0.000013 * [age] - 0.000639 * [heart])     (equation 3)

Standard Global Residual = ([Glob_Res] - 19.134585) / 35.306436     (equation 4)

where 19.134585 and 35.306436 are mean and standard deviation values of the global residual

LocRes = [y_1] - [yhat]     (equation 5)

StLocRes = ([Loc_Res] + 0.619157) / 19.353506     (equation 6)

where -0.619157 and 19.353506 are mean and standard deviation values of the global residual.

*Table 2. Global Regression Coefficient for each independent variable*

| Independent variable | Value of global regression coefficient $\beta_n$ |
|---|---|
| Intercept | $\beta_1$=2.067951 |
| TotPop | $\beta_2$= 0.000016 |
| Cancers | $\beta_3$= 0.000369 |
| 60-85yr | $\beta_4$= -0.000013 |
| Ischaemic_heart | $\beta_5$= -0.000639 |

## 3. Results

### 3.1 Global Regression Linear for global/aspatial model

The global regression model's results are presented in table 3, showing independent variables along with the global coefficient and their significance (Z-value). The global model results indicate Goodness-of-Fit ($R^2$) of 64 percent with an AICc value of 2060. The global regression model suggests a positive relationship between variable total Covid-19 and variables total population, total cancers, total kidney disease, and total asthma, while the rest are in a negative relationship.

*Table 3. Results of the global regression model, showing independent variables along with global coefficient, their significance, $R^2$, and AICc value*

| Variable | Global Regression | |
|---|---|---|
| | Estimated | Z-value |
| Intercept | 2.067951 | 62.750044 |
| TotPop | 0.000016 | 24.295802 |
| Cancers | 0.000369 | 13.438235 |
| 60-85yr | -0.000013 | -2.532562 |
| Ischaemic_heart | -0.000639 | -6.940932 |
| Resp_syst | -0.00047 | -9.340436 |
| Kidney | 0.000028 | 0.673674 |
| Asthma | 0.000703 | 5.041009 |
| **AICc** | | 2060.203967 |
| **Percent deviance explained (R Square)** | | 0.643084 |

### 3.2 Geographically Weighted Poisson Regression Linear for GWPR model

With reference to section 2.3.4 and the previous sections, only 4 variables are finally accepted for GWPR calibration. The Geographically Weighted Poisson Regression Linear (GWPR) model with an Adaptive bi-square Kernel resulted in an optimal bandwidth of 60 nearest neighbours. It means that a local model for each 130 local government area (LGA) was generated using weighted data from the nearest 60 LGA, which is 46 percent of the total LGA.

Local coefficients of the 4 variables for the entire NSW state are statistically presented in table 4 along with the AICc and $R^2$. For the purpose of interpretation, the estimated coefficients of individual variables and their significance (t-value) are mapped and presented in figures 3 – 8, including $R^2$ and standardized residuals maps. The GWPR model's result indicates Goodness-of-Fit ($R^2$) of 78% with an AICc value of 1264. Detailed discussion and interpretation of each map are presented in section 4.

*Table 4. Results of the GWPR model, showing independent variables along with local statistical coefficient, $R^2$ and AICc value*

| Variable | GWPR | | | |
|---|---|---|---|---|
| | Mean | STD | Min | Max |
| Intercept | 1.625614 | 1.360355 | -0.699802 | 3.013294 |
| TotPop | 0.000003 | 0.000011 | -0.000036 | 0.000012 |
| Cancers | 0.000808 | 0.000656 | 0.000399 | 0.002386 |
| 60-85yr | 0.000079 | 0.000169 | -0.000044 | 0.000442 |
| Ischaemic_heart | -0.002685 | 0.003282 | -0.009771 | -0.000357 |
| AICc | | | | 1264.086738 |
| Percent deviance explained (R Square) | | | | 0.785044 |

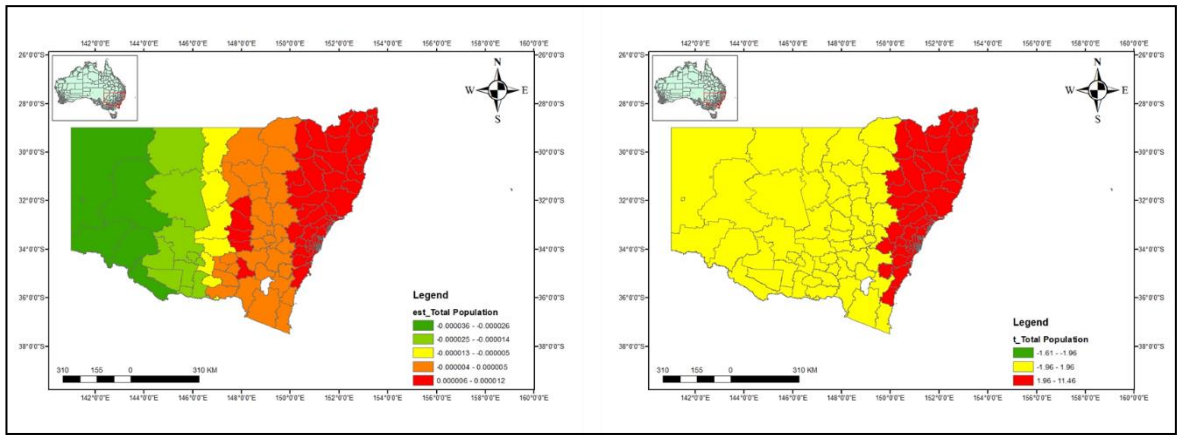

*Figure 3. Map of variable Total population, parameter estimate (left) and t-value (right)*

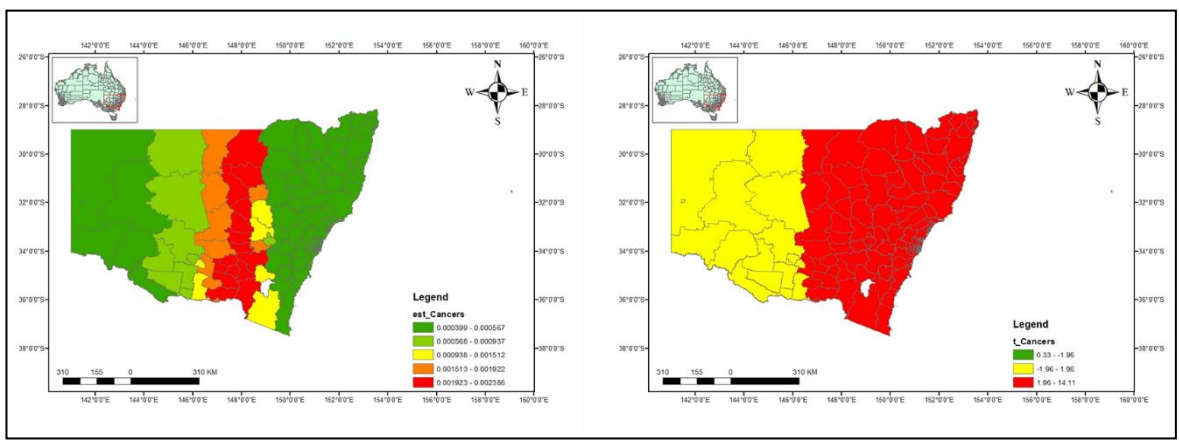

*Figure 4. Map of variable Cancers, parameter estimate (left) and t-value (right)*

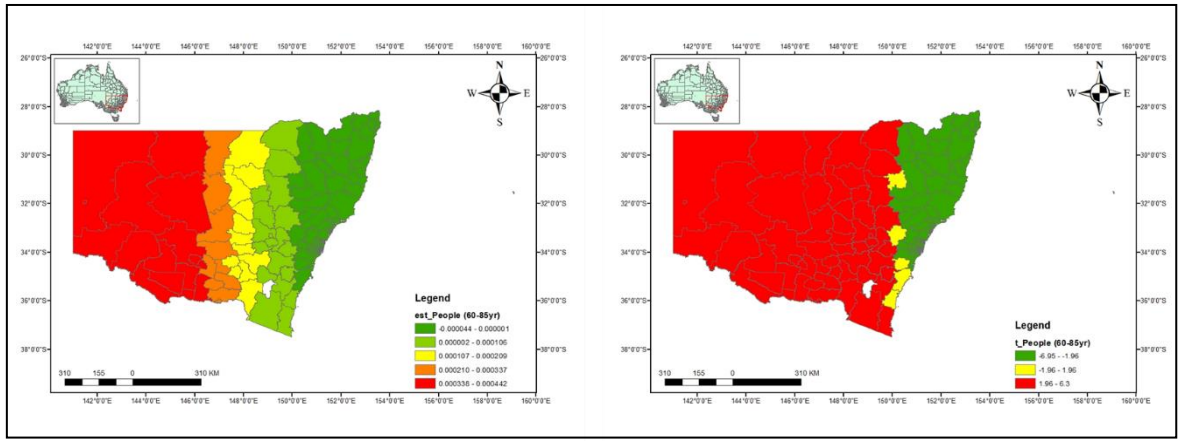

*Figure 5. Map of variable people with age between 60 and 85 years, parameter estimate (left) and t-value (right)*

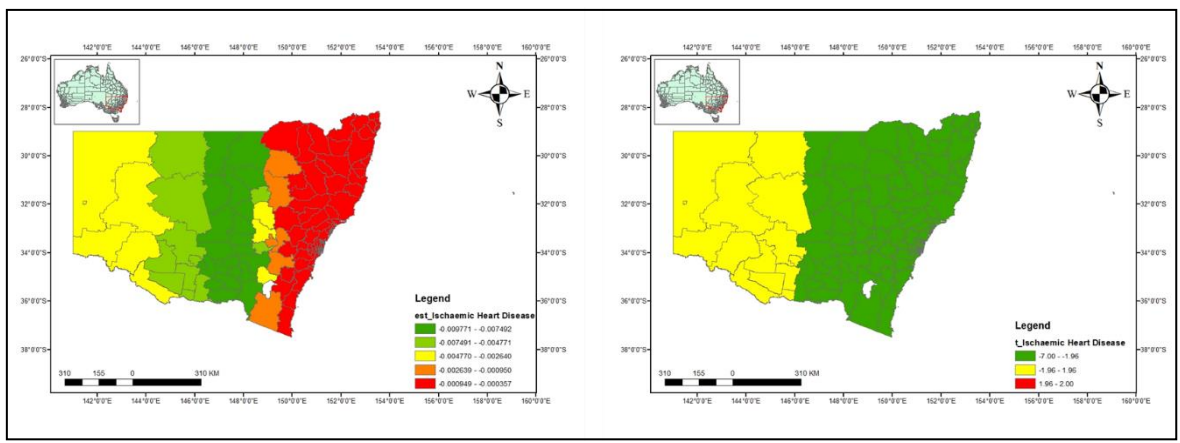

*Figure 6. Map of variable Ischaemic heart disease, parameter estimate (left) and t-value (right)*

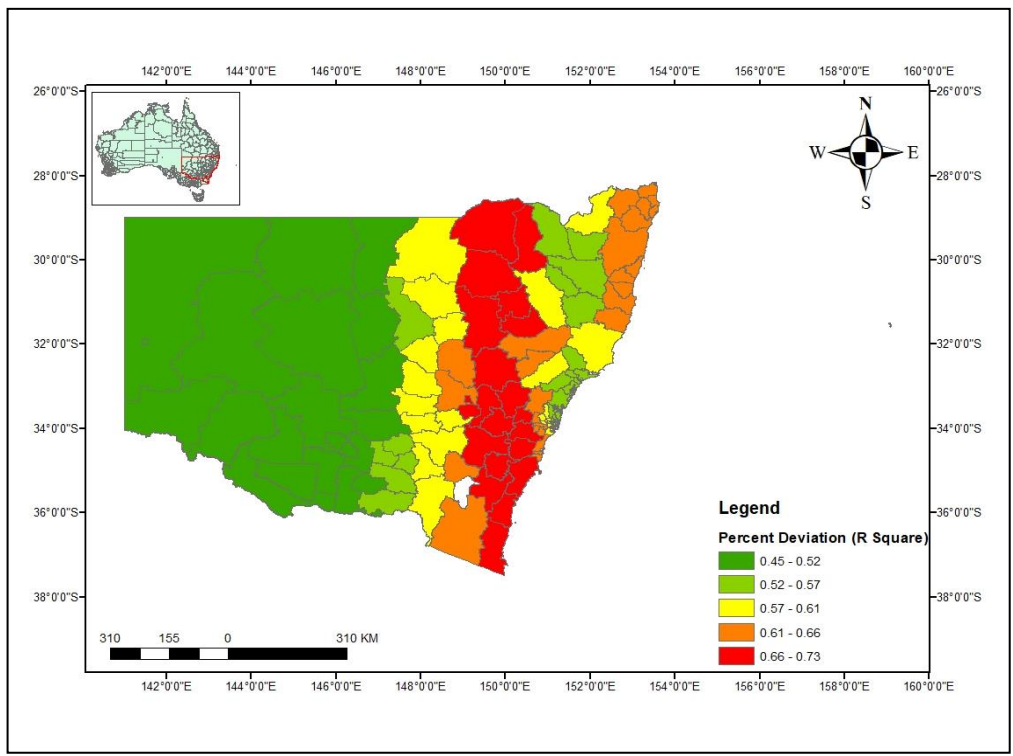

*Figure 7. Map of Goodness-of-Fit (Pseudo $R^2$)*

*Francelino A. Xavier-Conceição*

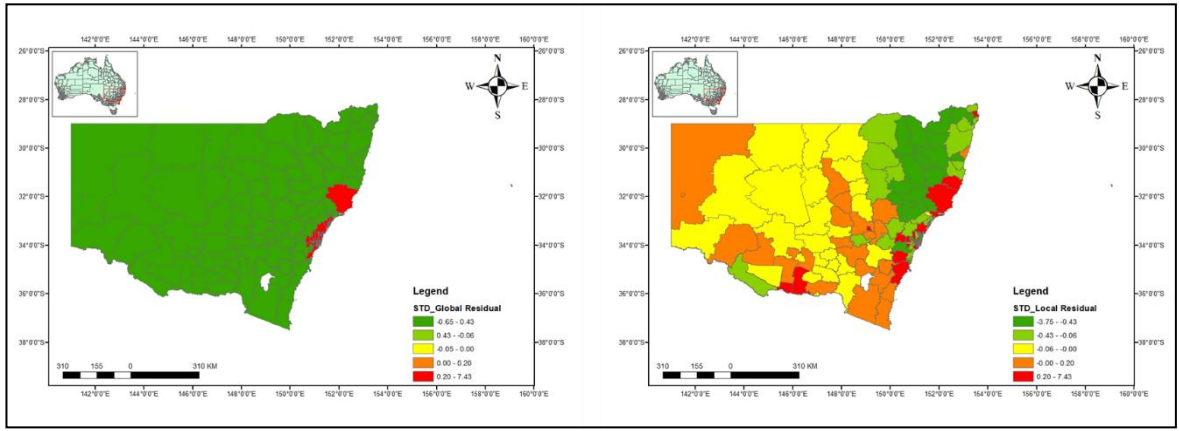

*Figure 8. Map of standardised global residual (left) and standardised local residual (right)*

## 4. Discussion

The GWPR model exhibits an improvement of model quality locally, where the AICc decreases from 2060 in the global model to 1264 in GWPR, and the Goodness-of-Fit ($R^2$) increases from 64% in the global model to 78% in GWPR.

Figures 3 – 6 show maps of the local parameter estimated (left maps), and the statistically significant to P<0.05 (right maps), which means the t-value between -1.96 and +1.96 are statistically not significant. The colour code for the local parameter (left maps) indicates the estimated risk level for Covid-19 infection, where the high risk is represented by red and decreases towards green. Meanwhile, the colour code for the t-value (right maps) indicates that only red and green colours are statistically significant and valid for this study.

Figure 3 illustrates the positive relationship between the variable total population and Covid-19 in the eastern side of NSW (red and orange in the left map), where the t-value is statistically significant (red in the right map). Similarly, figure 4 illustrates a positive relationship between variable cancers and Covid-19 in the eastern side of NSW (red, orange, yellow, and green in the left map), where the t-value is statistically significant (red in the right map). Both global and GWPR models suggest a positive relationship between these 2 variables and Covid-19.

Interestingly, the global model suggests a negative relationship between variables people with ages between 60 and 85 years and Covid-19. In contrast, the GWPR model presented in figure 5 reveals a positive relationship between these variables in most of the NSW area (red, orange, yellow, and light green in the left map), except the very eastern side of the NSW area with a negative relationship (green). The t-value is statistically significant in most of the study area, as indicated by green and red colours in the right map).

In contrast, the global model, as well as the GWPR model presented in figure 6, suggest a negative relationship between the variables ischaemic heart disease and Covid-19 in the eastern side of NSW (red, orange, light green, and green in the left map). The t-value for this relationship is statistically significant, as indicated by the green colour in the right map. The negative relationships between these variables observed from both global and local models are contradicted to the initial hypotheses by the WHO (2020). However, the estimated coefficients for this relationship are very small and close to zero; hence further observations including assessment from different perspectives are required for validation.

Figure 7 exhibits that the local model replicates variations in variable total Covid-19 very well in the eastern side of NSW with a Goodness-of-Fit ($R^2$) range between 52-73% but less fit in western NSW. Nevertheless, the $R^2$ in western NSW is not significantly poor and maintains its consistency between 45-52%; hence it can be accepted.

The standardized residuals for the global and local models (figure 8) indicate that the global model is spatially autocorrelated compared to the local model. The standardized residual of the local model (right map) suggests nonstationary relationships between the dependent variable total Covid-19 and the independent variables. Therefore, it proves the GWPR calibration's effectiveness to account for the geographical variability in this study.

## 5. Conclusions

The global regression and the Geographical Weighted Poisson Regression (GWPR) models are used in this study to assess the relationship between total Covid-19 and variables from socioeconomic and health sectors in local government area (LGA) in New South Wales (NSW). Through step-by-step procedure in building the global and GWPR model, 50 independent variables are filtered, and leaving 4 variables for the GWPR model calibration. The GWPR results exhibit that the independent variables total population, cancers, and people with ages between 60 and 85 years are positively correlated with dependent variable total Covid-19. Meanwhile, the independent variable ischaemic heart disease is negatively correlated with the dependent variable Covid-19, where the model coefficient is very small and close to zero; therefore, further assessment is required for validation.

The GWPR calibration successfully improved the model quality by 14%, and the AICc value decreased from 2060 in the global model to 1264 in GWPR. The Goodness-of-Fit ($R^2$) map indicates that the model performed well with $R^2$ varying from 45-73%. Furthermore, the standardized residual of the GWPR model exhibits nonstationary relationships between the dependent variable and independent variables compared to the global model. Therefore, GWPR calibration plays a vital role in geographic modelling at the local scale.

## 7. Annexes

**Annex 1. List of datasets available for this study. The variables accepted for model calibration are highlighted with green colour; and defined based on literature review and Multicollinearity check respectively.**

| Group | Variables | Description | Literature Review | Multicollinearity Check |
|---|---|---|---|---|
| Socioeconomic | SQKM | Square Kilometer of area per local government area (LGA) | Accepted | Accepted |
| | 0-19yr | Total population with age 0-19 years old | Accepted | Rejected |
| | 20-39yr | Total population with age 20-39 years old | Accepted | Rejected |
| | 40-59yr | Total population with age 40-59 years old | Accepted | Rejected |
| | 60-85yr | Total population with age 60-85 years old | Accepted | Accepted |
| | Lone_House | Total Lone person households | Rejected | Rejected |
| | Group_House | Total Group households | Rejected | Rejected |
| | Family_House | Total Family households | Rejected | Rejected |
| | Tot_households | Total households | Rejected | Rejected |
| | Couple_Children | Total Couple families with children under 15 and/or dependent students | Rejected | Rejected |
| | Couple_NoDep | Total Couple families with non-dependent children only | Rejected | Rejected |
| | Couple_NoChild | Total Couple families without children | Rejected | Rejected |
| | OnePar_Children | Total one parent families, children under 15 &/or dependent students | Rejected | Rejected |
| | onePar_NoDep | Total one parent families, non-dependent children only | Rejected | Rejected |
| | Tot_families | Total families | Rejected | Rejected |
| | Tot_mar_reg | Total married in a registered marriage | Rejected | Rejected |
| | Tot_mar_defact | Total married in a de facto marriage | Rejected | Rejected |
| | not_Marri | Total not married | Rejected | Rejected |
| | Tot_married | Total married | Rejected | Rejected |
| | never_Marri | Total never Married | Rejected | Rejected |
| | Tot_widowed | Total widowed | Rejected | Rejected |
| | Tot_divorced | Total divorced | Rejected | Rejected |
| | Tot_separated | Total separated | Rejected | Rejected |
| | TotPop | Total population_2016 | Accepted | Accepted |
| Health Records | Inf&Par_Dis | Total case of infectious and parasitic diseases in 2017 | Accepted | Rejected |
| | Cancers | Total case of cancers in 2017 | Accepted | Accepted |
| | End-nut-met | Total case of endocrine, nutritional and metabolic diseases in 2017 | Accepted | Rejected |
| | Diabetes | Total case of diabetes in 2017 | Accepted | Accepted |
| | Mental | Total case of mental health related conditions in 2017 | Accepted | Rejected |
| | Mood-Disord | Total case of mood affective disorders in 2017 | Accepted | Accepted |
| | Nervous_syst | Total case of nervous system diseases in 2017 | Accepted | Rejected |
| | eye-adnexa | Total case of eye and adnexa diseases in 2017 | Accepted | Rejected |
| | ear-mastoid | Total case of ear and mastoid process diseases in 2017 | Accepted | Rejected |
| | circ_syst | Total case of circulatory system diseases in 2017 | Accepted | Rejected |
| | Ischaemic_heart | Total case of ischaemic heart disease in 2017 | Accepted | Accepted |
| | Heart_fail | Total case of heart failure in 2017 | Accepted | Rejected |
| | Stroke | Total case of stroke in 2017 | Accepted | Accepted |
| | Resp_syst | Total case of respiratory system diseases in 2017 | Accepted | Accepted |
| | Asthma | Total case of asthma in 2017 | Accepted | Accepted |
| | Obst_Pulmonary | Total case of Chronic Obstructive Pulmonary Disease (COPD) in 2017 | Accepted | Accepted |
| | digestive_syst | Total case of digestive system diseases in 2017 | Accepted | Rejected |
| | skin-subcunt | Total case of skin and subcutaneous tissue diseases in 2017 | Accepted | Rejected |
| | Musc_syst | Total case of musculoskeletal system and connective tissue disease in 2017 | Accepted | Rejected |
| | Gen_syst | Total case of genitourinary system diseases in 2017 | Accepted | Rejected |
| | Kidney | Total case of chronic kidney disease in 2017 | Accepted | Accepted |
| | Perinatal_cond | Total case of certain conditions originating in the perinatal period in 2017 | Accepted | Rejected |
| | Birth defects | Total case of congenital malformations, deformations and chromosomal abnormalities in 2017 | Accepted | Rejected |
| | Poison-others_ext | Total case of injury, poisoning and other external causes in 2017 | Accepted | Rejected |
| | %smoke_Pregn | Total percent smoking during pregnancy | Rejected | Rejected |
| | Tot_Chil_Immun | Total children fully immunised | Rejected | Rejected |

**Annex 2. Matrix of Multicollinearity using Spearman's correlation rank (note: correlation values >0.7 are excluded except those considered as pre-condition for coronavirus infection).**

| | SQKM | TotPop | Inf&Par_Dis | Cancers | End-nut-met | Diabetes | Mental | Mood-Disord | Nervous_syst | eye-adnexa | ear-mastoid | circ_syst | Ischaemic_heart | Heart_fail | Stroke | Resp_syst | Asthma | Obst_Pulmonary | digestive_syst | skin-subcunt | Musc_syst | Gen_syst | Kidney | Perinatal_cond | Birth defects | Poison-others_ext | 0-19yr | 20-39yr | 40-59yr | 60-85yr |
|---|---|---|---|---|---|---|---|---|---|---|---|---|---|---|---|---|---|---|---|---|---|---|---|---|---|---|---|---|---|---|
| SQKM | 1.0 | | | | | | | | | | | | | | | | | | | | | | | | | | | | | |
| TotPop | -0.3 | 1.0 | | | | | | | | | | | | | | | | | | | | | | | | | | | | |
| Inf&Par_Dis | -0.3 | 1.0 | 1.0 | | | | | | | | | | | | | | | | | | | | | | | | | | | |
| Cancers | -0.3 | 0.9 | 0.9 | 1.0 | | | | | | | | | | | | | | | | | | | | | | | | | | |
| End-nut-met | -0.3 | 1.0 | 1.0 | 1.0 | 1.0 | | | | | | | | | | | | | | | | | | | | | | | | | |
| Diabetes | -0.3 | 0.9 | 1.0 | 0.9 | 0.9 | 1.0 | | | | | | | | | | | | | | | | | | | | | | | | |
| Mental | -0.3 | 1.0 | 0.9 | 0.9 | 0.9 | 0.9 | 1.0 | | | | | | | | | | | | | | | | | | | | | | | |
| Mood-Disord | -0.2 | 0.5 | 0.6 | 0.5 | 0.5 | 0.5 | 0.5 | 1.0 | | | | | | | | | | | | | | | | | | | | | | |
| Nervous_syst | -0.3 | 1.0 | 0.9 | 1.0 | 1.0 | 0.9 | 1.0 | 0.5 | 1.0 | | | | | | | | | | | | | | | | | | | | | |
| eye-adnexa | -0.3 | 0.9 | 0.9 | 1.0 | 0.9 | 0.9 | 0.9 | 0.4 | 0.9 | 1.0 | | | | | | | | | | | | | | | | | | | | |
| ear-mastoid | -0.3 | 0.9 | 0.9 | 1.0 | 1.0 | 0.9 | 0.9 | 0.5 | 1.0 | 0.9 | 1.0 | | | | | | | | | | | | | | | | | | | |
| circ_syst | -0.3 | 0.9 | 0.9 | 1.0 | 1.0 | 0.9 | 0.9 | 0.5 | 1.0 | 1.0 | 1.0 | 1.0 | | | | | | | | | | | | | | | | | | |
| Ischaemic_heart | -0.3 | 0.9 | 0.9 | 1.0 | 1.0 | 0.9 | 0.9 | 0.5 | 0.9 | 0.9 | 0.9 | 1.0 | 1.0 | | | | | | | | | | | | | | | | | |
| Heart_fail | -0.3 | 0.9 | 1.0 | 0.9 | 0.9 | 0.9 | 0.9 | 0.5 | 0.9 | 0.9 | 0.9 | 0.9 | 0.9 | 1.0 | | | | | | | | | | | | | | | | |
| Stroke | -0.3 | 0.9 | 0.9 | 1.0 | 0.9 | 0.9 | 0.9 | 0.4 | 0.9 | 1.0 | 0.9 | 1.0 | 0.9 | 0.9 | 1.0 | | | | | | | | | | | | | | | |
| Resp_syst | -0.3 | 1.0 | 1.0 | 0.9 | 1.0 | 0.9 | 0.9 | 0.5 | 0.9 | 0.9 | 1.0 | 1.0 | 0.9 | 1.0 | 0.9 | 1.0 | | | | | | | | | | | | | | |
| Asthma | -0.2 | 0.7 | 0.7 | 0.7 | 0.7 | 0.7 | 0.7 | 0.5 | 0.7 | 0.6 | 0.7 | 0.7 | 0.7 | 0.7 | 0.7 | 0.7 | 1.0 | | | | | | | | | | | | | |
| Obst_Pulmonary | -0.2 | 0.9 | 0.9 | 0.9 | 0.9 | 0.9 | 0.5 | 0.9 | 0.9 | 0.9 | 0.9 | 0.9 | 0.9 | 0.9 | 0.9 | 0.7 | 1.0 | | | | | | | | | | | | | |
| digestive_syst | -0.3 | 1.0 | 1.0 | 1.0 | 1.0 | 0.9 | 1.0 | 0.5 | 1.0 | 0.9 | 1.0 | 1.0 | 1.0 | 1.0 | 0.9 | 1.0 | 0.7 | 0.9 | 1.0 | | | | | | | | | | | |
| skin-subcunt | -0.3 | 1.0 | 1.0 | 0.9 | 1.0 | 0.9 | 0.9 | 0.5 | 1.0 | 0.9 | 1.0 | 0.9 | 0.9 | 0.9 | 0.9 | 1.0 | 0.7 | 0.9 | 1.0 | 1.0 | | | | | | | | | | |
| Musc_syst | -0.3 | 0.9 | 0.8 | 1.0 | 0.8 | 0.9 | 0.9 | 0.5 | 1.0 | 0.9 | 0.9 | 0.9 | 0.9 | 0.9 | 0.9 | 0.9 | 0.6 | 0.8 | 0.9 | 0.9 | 1.0 | | | | | | | | | |
| Gen_syst | -0.3 | 1.0 | 1.0 | 0.9 | 1.0 | 0.9 | 0.9 | 0.5 | 0.9 | 0.9 | 0.9 | 0.9 | 1.0 | 0.9 | 1.0 | 0.7 | 0.9 | 1.0 | 1.0 | 0.9 | 1.0 | | | | | | | | | |
| Kidney | -0.2 | 0.5 | 0.5 | 0.5 | 0.5 | 0.6 | 0.5 | 0.2 | 0.5 | 0.6 | 0.5 | 0.5 | 0.5 | 0.6 | 0.6 | 0.5 | 0.4 | 0.5 | 0.5 | 0.5 | 0.5 | 0.7 | 1.0 | | | | | | | |
| Perinatal_cond | -0.2 | 0.9 | 0.9 | 0.7 | 0.8 | 0.9 | 0.8 | 0.6 | 0.8 | 0.7 | 0.8 | 0.8 | 0.8 | 0.8 | 0.7 | 0.9 | 0.8 | 0.8 | 0.9 | 0.9 | 0.7 | 0.9 | 0.4 | 1.0 | | | | | | |
| Birth defects | -0.3 | 0.9 | 0.9 | 0.8 | 0.9 | 0.9 | 0.9 | 0.6 | 0.9 | 0.8 | 0.9 | 0.9 | 0.9 | 0.8 | 0.9 | 0.9 | 0.8 | 0.8 | 0.9 | 0.9 | 0.8 | 0.9 | 0.5 | 1.0 | 1.0 | | | | | |
| Poison-others_ext | -0.3 | 1.0 | 1.0 | 1.0 | 1.0 | 1.0 | 1.0 | 0.5 | 1.0 | 1.0 | 1.0 | 0.9 | 1.0 | 0.9 | 1.0 | 0.7 | 0.9 | 1.0 | 1.0 | 0.9 | 1.0 | 0.5 | 0.8 | 0.9 | 1.0 | | | | | |
| 0-19yr | -0.3 | 1.0 | 1.0 | 0.9 | 1.0 | 0.9 | 0.9 | 0.5 | 1.0 | 0.8 | 0.9 | 0.9 | 0.9 | 0.9 | 0.9 | 1.0 | 0.7 | 0.9 | 1.0 | 1.0 | 0.9 | 1.0 | 0.5 | 0.9 | 1.0 | 1.0 | 1.0 | | | |
| 20-39yr | -0.3 | 0.9 | 0.9 | 0.8 | 0.9 | 0.8 | 0.9 | 0.5 | 0.9 | 0.8 | 0.8 | 0.8 | 0.8 | 0.9 | 0.8 | 0.9 | 0.6 | 0.8 | 0.9 | 0.9 | 0.8 | 0.9 | 0.4 | 0.8 | 0.9 | 0.9 | 0.9 | 1.0 | | |
| 40-59yr | -0.3 | 1.0 | 1.0 | 0.9 | 1.0 | 0.9 | 1.0 | 0.5 | 1.0 | 0.9 | 1.0 | 0.9 | 0.9 | 0.9 | 0.9 | 1.0 | 0.7 | 0.9 | 1.0 | 1.0 | 0.9 | 1.0 | 0.5 | 0.9 | 0.9 | 1.0 | 1.0 | 0.9 | 1.0 | |
| 60-85yr | -0.3 | 1.0 | 0.9 | 1.0 | 1.0 | 0.9 | 0.9 | 0.5 | 1.0 | 1.0 | 0.9 | 1.0 | 1.0 | 1.0 | 1.0 | 1.0 | 0.7 | 0.9 | 1.0 | 1.0 | 0.9 | 1.0 | 0.5 | 0.8 | 0.9 | 1.0 | 0.9 | 0.8 | 1.0 | 1.0 |